\documentclass[aps,10pt,prb,reprint,twocolumn,showpacs,floatfix,superscriptaddress]{revtex4-2}

\usepackage{amsfonts,amsmath,amssymb}
\usepackage[margin=25.4mm,inner=25.4mm,marginpar=10mm]{geometry}
\usepackage{graphicx, epstopdf}
\usepackage{microtype}
\usepackage[figuresright]{rotating}
\usepackage{setspace}
\usepackage{textcomp}
\usepackage{times}
\usepackage[normalem]{ulem}
\usepackage[abs]{overpic}
\usepackage{color}

\usepackage[T1]{fontenc}
\usepackage[utf8]{inputenc}
\usepackage{physics}
\usepackage{bm}

\newlength{\figwidth}
\setcitestyle{super}

\begin{document}

\title{Single-, double-, and triple-slit diffraction of molecular matter-waves}

\author{Christian Brand}
\email{Christian.Brand@dlr.de}
\affiliation{University of Vienna, Faculty of Physics, Boltzmanngasse 5, A-1090 Vienna, Austria}
\affiliation{German Aerospace Center (DLR), Institute of Quantum Technologies, S\"oflinger Stra\ss e 100, 89077 Ulm, Germany}

\author{Stephan Troyer}
\affiliation{University of Vienna, Faculty of Physics, Boltzmanngasse 5, A-1090 Vienna, Austria}

\author{Christian Knobloch}
\affiliation{University of Vienna, Faculty of Physics, Boltzmanngasse 5, A-1090 Vienna, Austria}

\author{Ori Cheshnovsky}
\affiliation{Tel Aviv University, The Center for Nanosciences and Nanotechnology, Tel Aviv 69978, Israel}
\affiliation{Tel Aviv University, School of Chemistry, The Raymond and Beverly Faculty of Exact Sciences, Tel Aviv 69978, Israel}

\author{Markus Arndt}
\affiliation{University of Vienna, Faculty of Physics, Boltzmanngasse 5, A-1090 Vienna, Austria}

\date{\today}

\begin{abstract}
Even 100 years after its introduction by Louis de Broglie, the wave-nature of matter is often regarded as a mind-boggling phenomenon. 
To give an intuitive introduction to this field, we here discuss the diffraction of massive molecules through a single, double, and triple slit, as well as a nanomechanical grating. 
While the experiments are in good agreement with undergraduate textbook predictions, we also observe pronounced differences resulting from the molecules' mass and internal complexity.
The molecules' polarizability causes an attractive van der Waals interaction with the slit walls, which can be modified by rotating the nanomechanical mask with respect to the molecular beam. 
The text is meant to introduce students and teachers to the concepts of molecule diffraction, supported by problems and  solutions that can be discussed in class.
\end{abstract}

\maketitle

\section{Introduction}
Due to its conceptual simplicity, diffraction at single- and double-slits is often used to illustrate the principles of wave optics.
Extending this idea to matter waves then serves to illustrate the wave-particle duality and the superposition principle for massive particles in quantum mechanics.
The present work summarizes the analogies and differences of light and particle optics in the context of recent molecule diffraction experiments. 
We show that a proper description needs to include internal particle properties such as the polarizability, even though the de Broglie wavelength contains only information about the center of mass motion. 
  
Matter-wave diffraction through a few slits ($N=1-3$) has been demonstrated for electrons,\cite{Mollenstedt_ZPhys155_472, Jonsson_ZPhys161_454, Bach_NJP15_033018} neutrons,\cite{Shull_PhysRev179_752, Zeilinger_RevModPhys60_1067} atoms,\cite{Carnal_PRL66_2689, Shimizu_PRA46_R17,Shin_PRL92_050405,Szriftgiser_PRL77_4, Shin_PRL92_050405} and molecules,\cite{Nairz_PRA65_032109,Cotter_SciAdv3_e1607478} 
using mostly micro- or nanopatterned membranes.
The possibility to realize even complex structures with nanometer precision in these materials led to the fabrication of slits\cite{Carnal_PRL66_2689} and gratings,\cite{Keith_PRL61_1580} sieves\cite{Tennant_JVacSciTechnolB8_1975,Carnal_PRL67_3231} and holograms.~\cite{Fujita_Nature380_691}
They are used to focus beams of neutral helium\cite{Doak_PRL83_4229} and study weakly bound clusters\cite{Schoellkopf_Science266_1345, Grisenti_PRL85_2284} as well as higher-order matter-wave interference.\cite{Cotter_SciAdv3_e1607478,Barnea_PhysRevA97_023601}
Nanomechanical gratings became essential for three-grating interferometers with electrons,\cite{Cronin_PRA74_061602,Gronniger_NewJPhys8_224} neutrons,\cite{Zouw2000} atoms,\cite{Keith_PhysRevLett66_2693} and molecules\cite{Brezger_PhysRevLett88_100404,Gerlich_NatPhys3_711,Fein_NatPhys15_1242} up to masses beyond 25,000~u - and even for experiments with anti-matter.\cite{Sala_SciAdv5_eaav7610}

\begin{figure}
	\includegraphics[width=\linewidth]{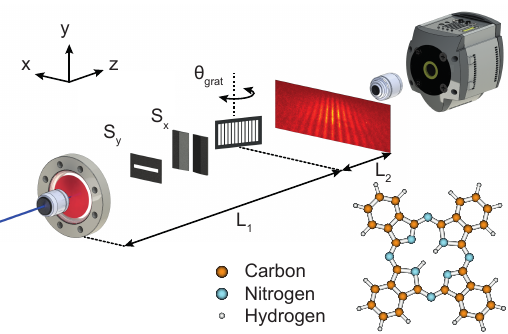}
	\caption{
	Laser-evaporation of molecules leads to a beam which is collimated along $x$ and $y$.
	After 1.55~m of free flight the matter-wave is diffracted at a nanomechanical grating, which can be rotated  by $\theta_{\rm{grat}}$. The resulting pattern is collected 0.59~m further downstream on a quartz plate and visualized via laser-induced fluorescence microscopy. We use the organic fluorophore phthalocyanine.
	}
	\label{fig:Setup}
\end{figure}

Here we put molecule diffraction at nanomechanical masks into a pedagogical context.
These experiments can serve in high school and undergraduate teaching, since they can be related directly to textbook knowledge of classical optics. 
Our work expands on an earlier discussion of fullerene diffraction in this journal in  2003,\cite{Nairz_AmJP71_319} explaining in more detail how to prepare and quantify transverse and longitudinal coherence of the molecular matter wave.
To visualize the wave-particle duality we use fluorescence imaging: Each molecule can be identified as a single particle on the detector and a large number of molecules then leads to the emergence of a diffraction pattern, as shown before for single photons\cite{Dimitrova_AmJPhys76_137} and electrons.\cite{Tonomura_AmJPhys57_117}

While diffraction of molecules and light share many fundamental features, we also see pronounced differences resulting from the particles' mass and electronic structure. Molecules fall visibly in the gravitational field and they interact with the grating walls because of van der Waals forces.
These forces are studied by rotating the grating relative to the molecular beam, which changes with the average distance between the molecules and the slit wall.
The role of the internal molecular structure opens numerous interesting questions, which are treated as problems and solutions for class work in the Appendix.

\section{Theory}
\subsection{Fraunhofer diffraction}
Like many textbooks, we discuss diffraction through a thin mask in the far-field approximation, that is, in the Fraunhofer regime. 
In this regime the curvature of the wavefronts can be neglected, and the diffracted intensity pattern is described by the Fourier transform of the transmission function of the diffracting mask (see Appendix~\ref{sec:SupplFourier}). 
Assuming an incident plane wave, the far-field assumption is valid when $L_2 \gg w^2/\lambda_{\rm dB}$, where $L_2$ is the distance between the mask and the detector, $w$ is the width of the coherently illuminated mask, and $\lambda_{\rm dB}$ the de Broglie wavelength (see Fig.~\ref{fig:Setup}). 
For $w=300$~nm and $\lambda_{\rm dB}\simeq 3\times 10^{-12}$~m, the transition region is $<50$~mm and thus considerably smaller than $L_2=590$~mm in our experiments.  
But even when this requirement is not met, the far-field may still be a reasonable approximation (see Appendix~\ref{Sec:SupplNear}).

The non-relativistic de Broglie wavelength $\lambda_{\rm dB}=h/mv$ depends on the particle's mass $m$ and velocity $v$ with Planck's constant $h$.
If we neglect all internal properties, the intensity distribution $I$ at an angle $\theta$ depends on the grating period $d$, the slit width $s$, and the number $N$ of coherently illuminated slits
\begin{equation}
I(\theta)\propto\underbrace{\left(\frac{\sin \beta}{\beta}\right)^2}_{\text{single slit}}\underbrace{\left(\frac{\sin N\alpha}{\sin\alpha}\right)^2}_\text{multiple slit} 
\label{eqn:Fraunhofer}
\end{equation}
with $\alpha=(\pi d/\lambda_{\rm dB})\sin\theta$ and $\beta=(\pi s/\lambda_{\rm dB})\sin\theta$.
The first term in Eq.~(\ref{eqn:Fraunhofer}) describes diffraction through a single slit while the second term accounts for $N$-slit interference.

\begin{figure}
	\includegraphics[width=\linewidth]{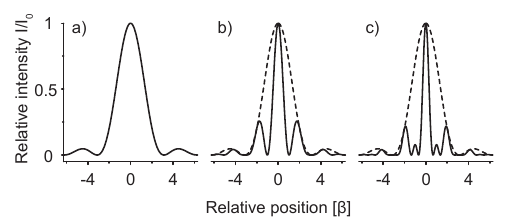}
	\caption{
Diffraction patterns through a single (a), a double (b), and a triple slit (c) according to Eq.~\ref{eqn:Fraunhofer}.
The envelope function is determined by smallest diffraction element, i.e. the  single slit, as indicated by the broken line for (b) and (c).
The $x$-axis is given in multiples of $\beta=(\pi s/\lambda)\sin\theta$ and the opening fraction (slit width/period) is 0.65.
	}
	\label{fig:slits}
\end{figure}

In quantum mechanics we can additionally arrive at a similar prediction for the width of the central diffraction lobe based on Heisenberg's uncertainty relation. 
When a wave traverses a slit, its position is confined to the width $s$. 
This induces a transverse momentum uncertainty with a full width at half maximum (FWHM) of
\begin{equation}
\Delta p = 0.89\,h/s \text{,}
\label{eqn:Heisenberg_slit}
\end{equation}
which evolves into a position uncertainty further downstream (see Appendix~\ref{sec:SupplCoherence}).
By measuring the beam's width, that is, its position uncertainty at the detector, we can thus extract $\Delta p$.\cite{Nairz_PRA65_032109,Shull_PhysRev179_752}
Within this envelope a multi-path interference pattern is formed when the wave is diffracted at two or more slits. 
Increasing the number $N$ of illuminated slits sharpens the principal interference fringes and causes the occurrence of $N-2$ secondary maxima. 
The expected diffraction patterns behind a single, a double, and a triple slit are shown in Fig.~\ref{fig:slits}.

\subsection{Coherence}

In order to observe multi-slit diffraction, the wavelets along different paths from the source to the detector need to have defined phase relations, i.e. they need to be coherent.
Mathematically, coherence is defined as the normalized correlation function of the waves in space and time.
There are several examples for coherent sources: lasers in optics or Bose-Einstein condensates in matter-wave science, where all photons or atoms are highly correlated over macroscopic distances in space and time.
But even thermal light and molecular beams can be prepared to be coherent.

It is often useful to distinguish between transverse and longitudinal coherence.
As for laser radiation, they describe the distance along the propagation direction (longitudinal) and perpendicular to it (transverse) over which the phase of the matter-wave is correlated.

Transverse coherence can be visualized as the uncertainty in the particle's transverse position, which makes it impossible to predict through which grating slit the particle will go.
Since both the transverse coherence function and the diffraction pattern behind a single slit are described by a Fourier transform, they share the same functional form: $(\sin(x)/x)^2$.\cite{Born_principles-of-optics}
If we define the transverse coherence width $X_T$ by the distance between the first order minima of this coherence function, we find $X_T  = 2 L_1 \lambda_{\rm dB}/s$, which grows inversely proportional with the source width $s$ (see Appendix~\ref{sec:SupplCoherence}).

Longitudinal coherence is a measure of spectral purity of the beam and thus depends on the distribution of molecular velocities. 
If the distribution is too broad, the interference pattern vanishes, as soon as the constructive interference of one wavelength overlaps with the destructive interference of another one. 
This definition leads to the respective coherence length $X_L =\lambda^2 /\Delta \lambda$. 
It differs by a factor of $1/2\pi$ from a definition based on propagating Gaussian wave packets.\cite{Adams_PhysRep240_143}
However, it has proven surprisingly useful in real-world experiments, which are not necessarily well described by Gaussian wave packets (see Appendix~\ref{sec:SupplCoherence}).

While the transverse coherence grows with the distance behind the source, this is not the case for the longitudinal coherence $X_L$ because the spectrum does not change in free flight. 
We can, however, increase $X_L$ by spectral filtering, that is, by selecting molecular velocities.

\section{Experimental Setup}
To record molecular diffraction patterns, we use the experimental setup sketched in Fig.~\ref{fig:Setup}.\cite{Juffmann_NatNanotechnol7_297}
A thin film of the molecule phthalocyanine (PcH$_2$, mass $m=514.5$~u) is coated onto a window, which is mounted onto a vacuum chamber at a base pressure of $P<1\times 10^{-7}$~mbar. 
To launch the matter wave, we focus a laser at $\lambda=420$~nm with a 50x objective onto the film. 
This is where the first quantum effect comes into play: 
The position of the emitted molecules is defined by the spot size $s_1$ of the focused laser beam, which is twice the laser waist $s_1=2 w_0 = (1.7 \pm 0.5)$~$\mu$m.
This localized evaporation can be seen as a position measurement.
Approximating the evaporation spot as a rectangle, we can use Eq.~(\ref{eqn:Heisenberg_slit}) to estimate  the associated transverse momentum uncertainty of $\Delta p\simeq 3.5\times 10^{-28}$~kg m/s.
When the molecules reach the grating after $L_1=1.55$~m, this has evolved into a position uncertainty. 
Heisenberg's principle thus helps us to prepare the transverse coherence required to
illuminate several slits by the same molecular wave function.
For PcH$_2$ moving at $v=250$~m/s the transverse coherence width amounts to $X_T\simeq 5.7$~$\mu$m, i.e. $57$ times the grating periods of $100$~nm.
\begin{figure}[bth]
	\includegraphics[width=\linewidth]{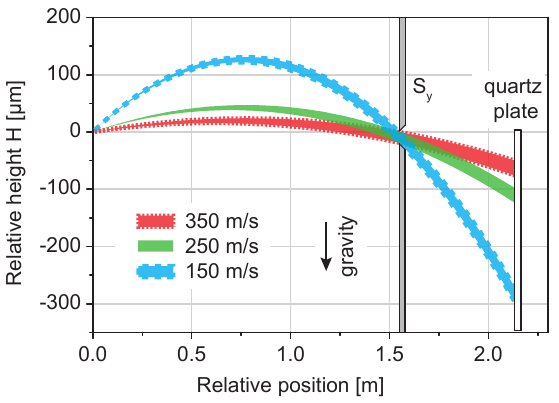}
	\caption{
The molecular beam spreads because of free-fall in the gravitational field. 
Molecules with a certain velocity pass the slit S$_y$ to fall to a certain height on the detector. 
The free-fall distance and the separation of the molecular velocities grow quadratically with time. The thickness of S$_y$ is not drawn to scale. 
	}
	\label{fig:velsel}
\end{figure}

The grating was milled with a focused beam of gallium ions into an ultra-thin membrane of amorphous carbon with a thickness of $T=21\pm 2$~nm. 
A 4-axis manipulator allows for 3D translations of the grating and a rotation around the $y$-axis.
In our ($N=1-3$)-slit gratings, the mean geometrical slit width amounts to $s_{\rm{geo}}=80\pm 5$~nm and the period is $d_{\rm{geo}}=100\pm 5$~nm.\cite{Cotter_SciAdv3_e1607478}
The bars between the slits have a cross section of $21\times 20$~nm$^2$, only an order of magnitude wider than the diameter of the diffracted molecule itself (1.5~nm). 
In the rotation experiments, we use a grating with $d_{\rm{geo}}=101\pm 2$~nm and $s_{\rm{geo}}=61\pm 1$~nm.
The maximum width of the patterned area amounts to $5$~$\mu$m and thus acts as a collimator in the $x$-direction. 
Additionally, the beam is loosely collimated by a piezo-controlled slit (S$_x$) to prevent transmission through membrane defects.

After passing the grating, the molecules are collected on a quartz plate $L_2=0.59$~m further downstream.
To visualize the molecular density pattern we excite PcH$_2$ with 60~mW of laser light at 661~nm focused onto a spot size of $400\times 400$~$\mu$m$^2$ and collect the laser-induced fluorescence with a 20x microscope objective.
A band filter transmitting in the wavelength region from 700 to 725~nm is used to separate the fluorescence from the laser light, and the image is recorded with an electron multiplying (EM) CCD camera.

In the experiments we use phthalocyanine (cf. Fig.~\ref{fig:Setup}) due to its high thermal stability and fluorescence quantum yield. 
This allows detection of single molecules  with high contrast, which is required to visualize even weak signals.\cite{Juffmann_NatNanotechnol7_297}
Furthermore, PcH$_2$ is non-polar and therefore essentially unperturbed by residual charges in the mechanical masks.\cite{Knobloch_FortschrPhys65_1600025}

We use a thermal beam, containing a wide range of molecular velocities and thus de Broglie wavelenghts.
If all these velocities were to overlap at the detection screen, it would obscure many of the details present in the diffraction pattern.
To prevent this, we restrict different velocities to different regions of the quartz plate.
For that purpose, we use a horizontal  delimiter S$_y$ (see Fig.~\ref{fig:velsel}). 
In combination with the source, it defines the free-flight parabolas of the molecules in the presence of gravitational acceleration $g$.
As the free-fall distance $H=g(L_1+L_2)^2/(2v^2)$ depends on the molecular velocity $v$, slow molecules fall further than fast ones and thus they are separated at the position of the detector.
In our few-slit diffraction experiment the masks themselves serve as velocity selectors, for the case of the diffraction grating, an additional slit is introduced just before it.

\begin{figure*}
	\includegraphics[width=0.9\linewidth]{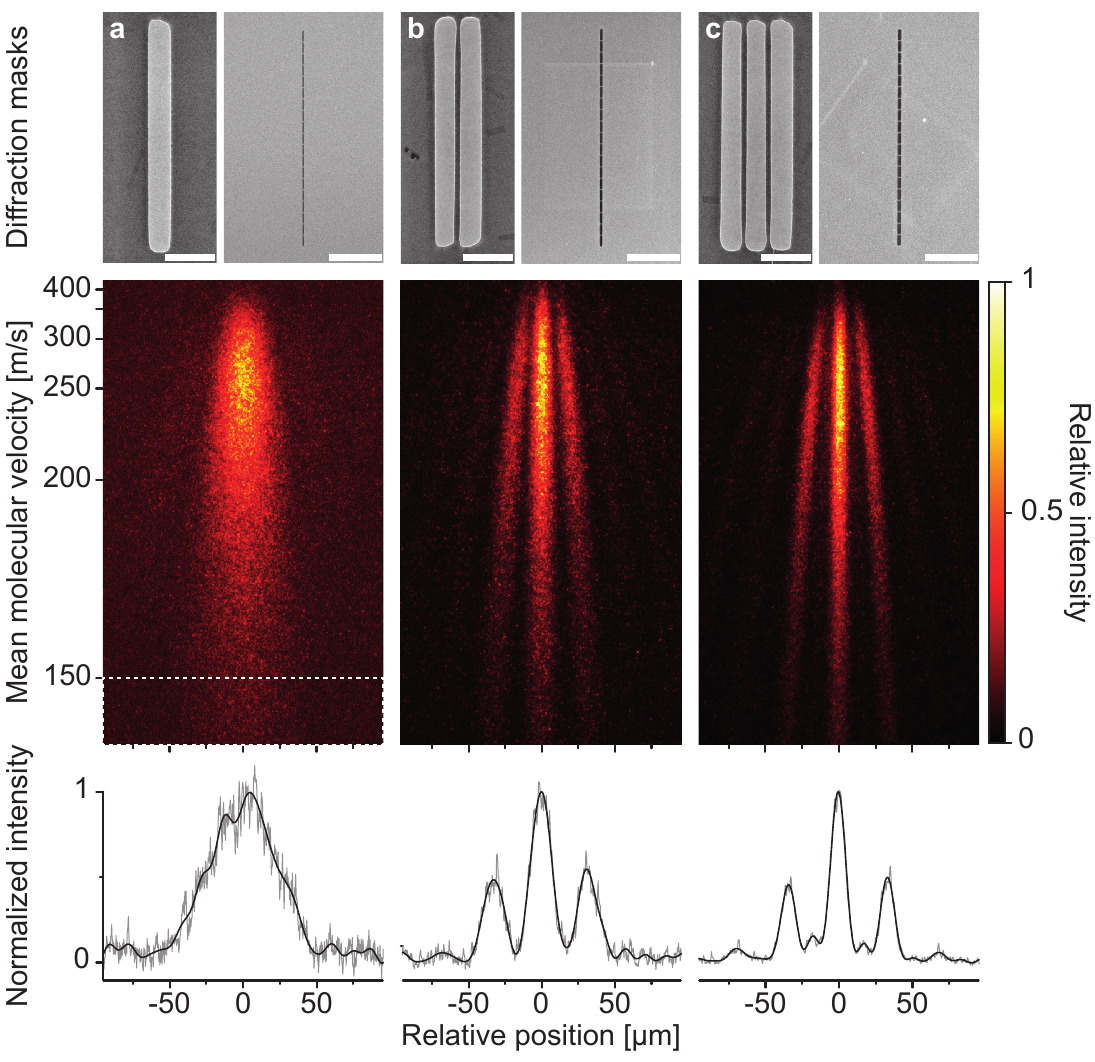}
	\caption{
Electron micrographs of a single (a), a double (b), and a triple slit (c).\cite{Cotter_SciAdv3_e1607478}
Each mask consists of $19$ units of the original pattern placed above each other to increase the molecular flux.
The slit width amounts to $80\pm 5$~nm and the period is $100\pm 5$~nm.
The scale bars correspond to 200~nm (left) and 5~$\mu$m (right), respectively. Sending molecular matter-waves through the masks results in diffraction patterns spanning the range from $140$ to $430$~m/s.
The lower trace shows the vertical sum over the velocity band from $140$ to $150$~m/s, indicated by the broken line in the signal of the single-slit pattern. 
The summed traces are overlaid with a low pass filter with a spatial cut-off frequency of $1/12.5$~$\mu$m.
	}
	\label{fig:Overview}
\end{figure*}

\section{Diffraction through a few slits}
\subsection{Results}

In Fig.~\ref{fig:Overview} we show the results of molecular matter-wave diffraction at a single, a double, and a triple slit. 
The electron micrographs of the respective masks are shown in the upper trace, and the molecular diffraction patterns are shown in the middle trace.
They span molecular velocities $v$ between $140$ and $430$~m/s, corresponding to de Broglie wavelengths $\lambda_{\rm{dB}} $ between $5.5$ and $1.8$~pm.
Each bright dot corresponds to a single molecule that scatters thousands of fluorescence photons during detection. 
At first glance we observe a qualitative difference between the single-slit pattern and the other two: Diffraction at a single slit leads to a structureless, broad signal while the double- and triple-slit pattern exhibit a sub-structure. This is what we expect based on Eq.~(\ref{eqn:Fraunhofer}) and Fig.~\ref{fig:slits}.

To illustrate the level of detail in the patterns, we vertically sum over the velocity band between $140$ and $150$~m/s as shown in the lower trace of Fig.~\ref{fig:Overview}.
Here all patterns share the same envelope resulting from single-slit diffraction.
The triple slit pattern also reveals the expected secondary maximum in
between the principal diffraction orders.
Furthermore, the width of the zeroth diffraction order decreases with increasing $N$ from $15.6\pm 0.1$~$\mu$m for the double slit to $12.0\pm 0.1$~$\mu$m for the triple slit.

It is often stated in textbooks that the wavelength has to be comparable to the grating period to observe interference. 
However, in the current experiments $\lambda_{\rm{dB}}$ is five orders of magnitude smaller than $d$, and still we observe high-contrast interference.
To achieve this, we have to fulfill three conditions:

First, the transverse coherence has to be large enough to engulf all slits that shall contribute to multi-slit interference.
As discussed before, $X_T$ covers dozens of grating periods and thus exceeds the minimum requirement by far.

Second, the collimation angle  has to be smaller than the diffraction angle $\theta=\lambda_{\rm{dB}}/d$ to prevent the diffraction orders from overlapping.
In our experiment the width of the source ($s_1=1.7$~$\mu$m) and the grating ($w=280$~nm for the triple slit) at a distance of $L_1=1.55$~m define a collimation angle of $\theta=(w+s_1)/L_1=1.3$~$\mu$rad.
This is well below the diffraction angle of $31$~$\mu$rad for PcH$_2$ moving at $250$~m/s.

Finally, the resolution of the detector has to be sufficient to resolve the pattern. 
This is achieved using fluorescence microscopy:
Each pixel in Fig.~\ref{fig:Overview} images $400\times 400$~nm$^2$ on the fluorescence screen.\cite{Juffmann_NatNanotechnol7_297}
Curve fitting allows to determine the barycenter of the fluorescence curve with 10~nm accuracy and hence can easily resolve features on the $\mu$m-scale, as observed here.

\subsection{Differences between light and matter-waves}

The patterns in Fig.~\ref{fig:Overview} display many features also expected for light: the quantitative separation of the principal diffraction peaks, the emergence of the intermediate peaks for $N=3$, and the narrowing of the principal diffraction orders with increasing $N$.
However, there are also some major differences.
The most outstanding feature is that the diffraction orders are not parallel but bent.
Our source emits a distribution of molecular velocities, which is reflected in the spread of de Broglie wavelengths $\lambda_{\rm{dB}}$.
As the molecules follow free-flight parabolas in the gravitational field, all diffraction orders except the zeroth should be curved, as can be seen in Fig.~\ref{fig:Overview}.

The narrowing of the diffraction orders with increasing $N$ is limited in our experiments.
In the model of Eq.~(\ref{eqn:Fraunhofer}), the diffraction orders get sharper as long as $N$ increases.
In the experiments, however, the width of the diffraction orders has a lower bound defined by the transverse collimation angle, i.e. deviations from the assumption of an incident plane wave.

Also the envelope function hints at differences in the diffraction mechanism.
From the micrographs we extract a geometrical slit width $s_{\rm{geo}}=80\pm 5$~nm.
According to Eq.~(\ref{eqn:Heisenberg_slit}), this should lead to a diffraction envelope with a FWHM of $35\pm 2$~$\mu$m at the detector for $v=145$~m/s.
However, we observe a width of $53\pm 1$~$\mu$m, corresponding to an effective slit width $s_{\rm{eff}}$ of $53\pm 1$~nm.
The reason for this reduction is the van der Waals, or more generally, the Casimir-Polder interaction.

\section{Van der Waals interactions}

\begin{figure}
	\includegraphics[width=\linewidth]{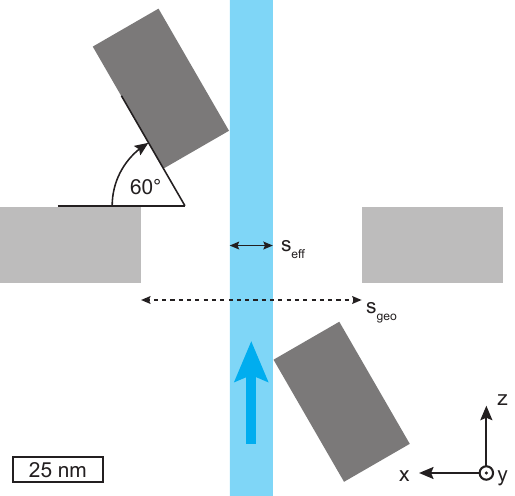}
	\caption{
Rotating the grating (light grey) by $60^\circ$ (dark grey) reduces the projection of the geometrical slit width $s_{\rm{geo}}$ onto the molecular beam to $s_{\rm{eff}}$ (blue area).
The molecules propagate in the direction of the blue arrow.
	}
	\label{fig:rotgrat}
\end{figure}

Casimir-Polder interactions result from fluctuations in the electron density of nearby objects.\cite{Casimir_PhysRev73_360}
As the maximum distance between molecule and the nearest grating bar is 40~nm during transmission, we are in the short range limit, known as the van der Waals interactions.
Here the induced dipole moments interact with their mirror images in the material.
While the attractive potential scales with $1/x^6$ between two isolated particles, we have to integrate over the half-sphere of the grating, resulting in a potential which scales with $1/x^3$.\cite{Mullins_AerosolSciTechnol17_105,Lennard-Jones_TransFaradaySoc28_333}
Approximating the grating thickness $T$ as infinite, the potential $V_{\rm{pot}}$ between the molecules and the grating can be written in the form $V_{\rm{pot}}(x)=-C_3/x^3$.
The factor $C_3$ includes the frequency-dependent polarizability of the particle and the dielectric function of the material grating, i.e. the response of both interaction partners to oscillating electric fields.
This leads to a position-dependent phase $\phi$, which is imprinted onto the molecular matter wave in each slit:
\begin{equation}
\phi\simeq\exp(-\frac{i}{\hbar v}\int_0^T{\frac{C_3}{(s/2-x)^3}+\frac{C_3}{(s/2+x)^3}dz}).
\label{eqn_phase}
\end{equation}
The potential strongly depends on the distance $s/2\pm x$ from the molecule to the grating walls and we have to consider the influence of both walls in the slit, leading to a double-sided potential. 
The expression in Eq.~(\ref{eqn_phase}) is justified for gratings whose thickness is about $10^4$ times the size of the diffracted particles.\cite{Grisenti_PhysRevLett83_1755}
For ultra-thin gratings, however, the molecule can actually be thicker than the grating itself.\cite{Brand_NatNanotechnol10_845}
Thus, the approach to the grating and the departure from it also have to be considered.\cite{Brand_AnnPhys(Berlin)527_580}

To fully describe the interaction we would have to characterize the molecule, the grating, and the interaction between them to a high level.\cite{Brand_AnnPhys(Berlin)527_580,Fiedler_AnnPhys(Berlin)527_570}
Such an analysis is very demanding: 
Each molecule consisting of $n$ atoms has $3n-6$ vibrations and many (often up to 500) rotational levels excited.
Flexible molecules can adopt a number of different conformations, which may interconvert within picoseconds.
And even for rigid molecules such as PcH$_2$, the polarizability is often not isotropic.
In consequence, the force depends on the orientation of the molecule during the transit through the grating.
Moreover, charges implanted in the grating material may lead to an attractive force several times stronger than expected.\cite{Brand_AnnPhys(Berlin)527_580}

Here, we resort to a phenomenological analysis:
Close to the grating walls, the phase shift becomes so large that small position changes cause large phase fluctuations and the interference terms are averaged out.
In a few nanometer distance the molecule may even be adsorbed by the grating.\cite{Juffmann_FoundPhys42_98}
Hence, we divide the slit into two regions:
In the center $\phi$ is small and multi-slit diffraction is possible.
Close to the grating walls, however, the molecule cannot contribute to the diffraction pattern any more.
In this picture the influence of the van der Waals interactions reduces the slit width from $s_{\rm{geo}}$ to an effective slit width $s_{\rm{eff}}$.\cite{Grisenti_PhysRevLett83_1755}

There are several options to modify the van der Waals interactions.
First, altering the grating material changes the coefficient $C_3$, which determines the phase shift.
Second, one can minimize the grating's thickness $T$ - ultimately to just a single layer of atoms.
This has recently been demonstrated using patterned single-layer graphene, which was stable enough to withstand the impact of fast molecules, and to yield high contrast molecule diffraction patterns.\cite{Brand_NatNanotechnol10_845}
Finally, we can change the molecule-grating distance and interaction length by rotating the grating.

\section{Diffraction through a rotated grating}

Rotating the grating modifies not only the effective grating period $d_{\rm{eff}}$ and slit width $s_{\rm{eff}}$, but also the interaction time and distance between the molecule and the grating walls.
The molecules get close only to the edges of the grating, as shown in Fig.~\ref{fig:rotgrat}.
This effect has been used to characterize nanomechanical gratings,\cite{Cronin_PRA70_043607,Grisenti_PRA61_033608} and the ensuing reduction in slit width was key to study weakly bound clusters.\cite{Bruhl_PRL95_063002}
In these experiments the maximum angle of $42^\circ$ was limited by the membrane thickness.\cite{Grisenti_PRA61_033608}
Here we use an ultra-thin grating with a thickness of $T=21\pm 2$~nm and a large opening fraction to achieve rotation angles up to $60^\circ$. 
This reduces the effective period by a factor of $\cos(60^\circ) =0.5$.

\subsection{Results}

\begin{figure}[tbh]
	\includegraphics[width=\linewidth]{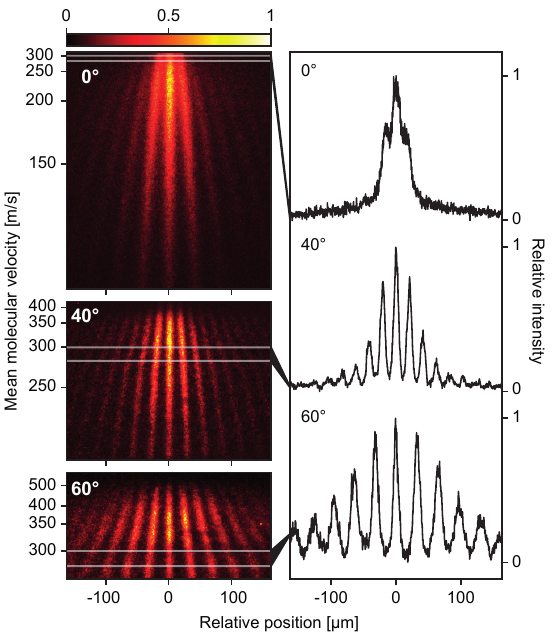}
	\caption{
	Diffraction patterns for PcH$_2$ diffracted at the grating rotated by $0^\circ$ (top), $40^\circ$ (middle), and $60^\circ$ (bottom).
	Rotating the grating reduces $d_{\rm{eff}}$ and $s_{\rm{eff}}$ and thus  increases the separation of the peaks and widens the single-slit envelope.
	The traces in the right column show the vertical sum over the velocity band from $280$ to $300$~m/s.
	To determine $s_{\rm{eff}}$ the maxima of the peaks are fitted with a Gaussian.
	The FWHM of the Gaussian envelope amounts to $39\pm 1~\mu$m ($0^\circ$), $69\pm 1~\mu$m ($40^\circ$), and $184\pm 2~\mu$m ($60^\circ$). 
	Each velocity axis has been scaled individually for best legibility.
	}
	\label{fig:RotGrat}
\end{figure}

The diffraction patterns recorded at $\theta_{\rm{grat}} =0^\circ$, $40^\circ$, and $60^\circ$ are shown in Fig.~\ref{fig:RotGrat}.
They span molecular  velocities $v$ from $500$ to $110$~m/s, corresponding to de Broglie wavelengths $\lambda_{\rm{dB}}$ between 1.6 and 7.1~pm.
For $\theta_{\rm{grat}} =0^\circ$ the pattern is dominated by the zeroth and both first diffraction orders.
Rotating the grating broadens the envelope function and shifts the position of the diffraction orders.
For $\theta_{\rm{grat}}=40^\circ$ the effective grating period $d_{\rm{eff}}=d_{\rm geo} \cos({\theta_{\rm{grat}}})$ is reduced to 77~nm, resulting in larger diffraction angles.
Due to the grating thickness the slit width is reduced from $s_{\rm{geo}}=61$~nm to $s_{\rm{eff}}=32$~nm.
This confinement of the matter wave in the effective slit widens the single-slit envelope and leads to a stronger population of higher diffraction orders.
In consequence, diffraction up to the $\pm 6^{\rm{th}}$ orders can be observed.
At $60^\circ$ the effective period $d_{\rm{eff}}$ is half of the geometrical one, and the slit width is reduced by a factor of 5.
At this angle, the diffraction envelope is about five times wider than under normal incidence.
 
We show typical traces for $v\sim 290$~m/s in the right column of Fig.~\ref{fig:RotGrat}.
To assess $s_{\rm{eff}}$, we fit a Gaussian to the maxima of the diffraction orders  and convert its FWHM to the corresponding slit width utilizing Eq.~(\ref{eqn:Heisenberg_slit}).
The width of the signal increases from $39\pm 1~\mu$m ($0^\circ$)  to $184\pm 2~\mu$m ($60^\circ$), corresponding to a decrease in $s_{\rm{eff}}$ from $36\pm 1$ to $8\pm 1$~nm.
Comparing the effective slit widths to the geometrical (Tab.~\ref{tab:reduction}) shows that the difference between them decreases with larger rotation angle:
While the difference amounts to $25\pm 2$~nm at perpendicular angle of incidence, it is reduced to only $4\pm 3$~nm at $60^\circ$.
\begin{table}
\caption{
Comparison of the geometry parameter for a mean velocity of 290~m/s. 
The difference $\Delta$ between the geometrical slit width $s_{\rm{geo}}$ and the effective one $s_{\rm{eff}}$ decreases considerably when rotating the grating from $0^\circ$ to $60^\circ$.
}
\noindent
\centering
\begin{tabular}{r|rrrrrr}
\hline\hline
																																			& &$0^\circ$		& & $40^\circ$		& &	$60^\circ$			\\
\hline
$s_{\rm{geo}}$ [nm]																			& &$61\pm 1$	& & $32\pm 2$	& & $12\pm 2$		\\
FWHM [$\mu$m]																		& &$39\pm 1$	& &	$69\pm 1$	& & $184\pm 2$	\\
$s_{\rm{eff}}$ [nm]																			& &$36\pm 1$	& & $20\pm 1$	& & $ 8 \pm 1$		\\
$\Delta (s_{\rm{geo}}-s_{\rm{eff}})$	[nm]	& &$25\pm 2$	& & $12\pm 3$ 	& & $  4\pm 3$		\\
\hline\hline
\end{tabular}
\label{tab:reduction}
\end{table}
However, such values have to be treated with care.
Prior experiments have shown that the width of the envelope may be smaller than expected for slit widths of a few nanometer.\cite{Grisenti_PRA61_033608}
In consequence, $s_{\rm eff}$ extracted for $60^\circ$ represents an upper bound.

\section{Summary and Outlook}

We have demonstrated molecular diffraction at a single-, double-, and triple slit as well as a rotated nanomechanical grating. 
Within the framework of the de Broglie hypothesis, the patterns agree astonishingly well with predictions from general wave optics, as used for light. 
However, we also observe pronounced differences, associated with the molecular mass and complex internal dynamics: Molecules fall visibly in the gravitational field and they are attracted by nearby walls. 
In the experiments presented here, the de Broglie wavelength ranges between $2-6$~pm. 
Even though it is smaller than each molecule by about three orders of magnitude, we can see diffraction and a high-contrast interference pattern: 
Grating diffraction thus translates a relative path length difference of a few picometers into peak separations on the order of dozens of micrometers.
This is a magnification by more than $10^6$.
In this respect, our experiments resemble small-angle X-ray scattering, aimed to reveal long range order in bio-systems.
We have shown how this magnification can be utilized to visualize the force due to the van der Waals interaction, which is here on the atto-newton level.
It is interesting to see that matter-wave based quantum technologies - using full multi-grating interferometers -- have started to generate impact in force and acceleration sensing applications.\cite{Arndt_PhysToday67_30,Tino2014,Schleich_ApplPhysB122_130}

Matter-wave diffraction requires delicate setups, making it challenging for students to gain hands-on experience.
A good alternative are online simulators, which provide a detailed lab environment and offer the possibility to perform realistic experiments life in class.\cite{quantuminteractive,Tomandl_SciRep5_14108}

The similarities and differences in the diffraction of matter and light are a good starting point to introduce matter-wave diffraction in introductory classes on quantum physics. 
To facilitate this, we include in the Appendix a number of problems and solutions related to our present experiments.

\section*{Acknowledgements}

We thank Thomas Juffmann and Joseph Cotter for work on that experiment as well as Yigal Lilach for writing the masks.
This project has received funding from the Austrian Science Fund (FWF) within project P-30176.

\renewcommand{\figurename}{Suppl. Fig.}
\setcounter{figure}{0}

\clearpage

\appendix

\section{The transition from near- to far-field}
\label{Sec:SupplNear}

We start with a short reminder of the transition between near- and far-field optics.
This holds some surprises that are less frequently discussed in textbooks even though they provide insight into fundamental optics phenomena and allow to derive practical predictions for real-world experiments. 
Diffraction is often treated as a \textit{far-field} phenomenon, where the dynamics can be approximated by the propagation of plane waves and effects of wave front curvature are neglected.
In the following we analyze when this is actually a good approximation.

\begin{figure}[b]
	\includegraphics[width=\linewidth]{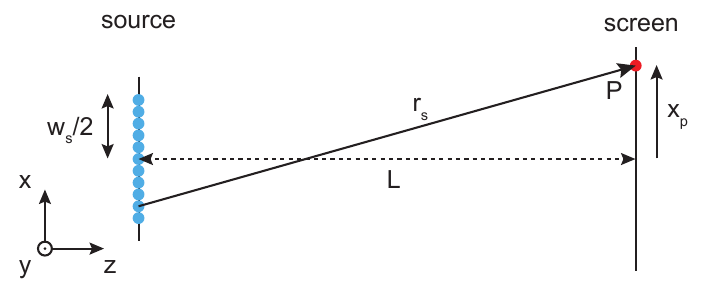}
	\caption{
	Each source point (blue) adds with a position-dependent phase $\phi_s$ to the image at point $P$ (red) on the screen.
	}
	\label{fig:NahundFern}
\end{figure}

Consider a source $s$ of width $w_s$ emitting monochromatic radiation of wavelength $\lambda$. 
If we divide $s$ into a sequence of point emitters positioned at $x_s$, each of them contributes with a path-length dependent phase $\phi_s=2\pi r_s/\lambda$ to the image at a point $P$ on a screen at distance $L$ (cf. Suppl.~Fig.~\ref{fig:NahundFern}).
We can expand the influence of the distance to $P$ to second order in $x$ according to: 
\begin{align}
    r_s &=\sqrt{L^2+(x_p-x_s)^2} \\
        &\simeq L+\frac{(x_p-x_s)^2}{2L}-\cdots \nonumber  \\
        &\simeq L+\frac{x_p^2}{2L}-\frac{x_p x_s}{L}+\frac{x_s^2}{2L}~. \nonumber
\end{align}
The curvature of the wave fronts can be ignored if the contribution of the quadratic terms to the phase is negligible.
If we consider only the source with $|x_s| \leq w_s/2$, we get
\begin{equation}
    \phi(\text{quad.})\simeq\frac{\pi w_s^2}{4\lambda L}\ll 1~.
\end{equation}
This expression imposes a condition for the minimal distance between grating and source or detector for the far-field approximation to be valid: 
\begin{equation}
    L_{\mathrm{NearToFar}}\gg  \pi w_s^2/4 \lambda.
    \label{eqn_neartofar}
\end{equation}

In our experiments the width of the nanomechanical grating is $w_s=5$~µm.
For molecules with a de Broglie wavelength of $\lambda_{\rm dB}=3\times 10^{-12}$~m the transition to the far field is expected at $8.3\,$m -- which exceeds both the source-to-grating distance $L_1=1.55$~m and the grating-to-detector distance $L_2=0.59$~m by about an order of magnitude. 
And yet, we observe an interference pattern (Fig.~6) that is well described by the textbook far-field formula.
So, Eq.~(\ref{eqn_neartofar}) seems to contradict our experiments.

\begin{figure}[tbh]
	\includegraphics[width=\linewidth]{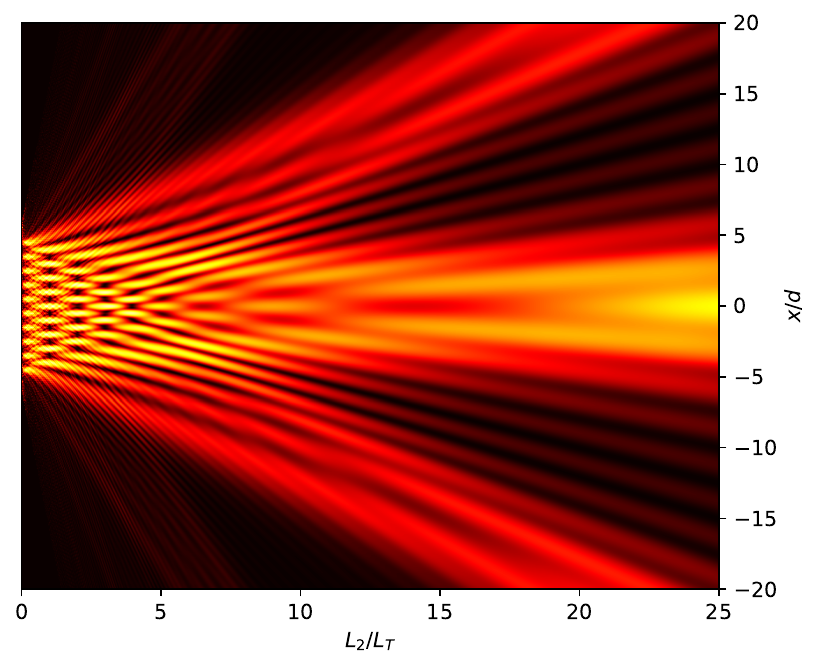}
	\caption{Transition from near-field to far-field interference for $N=10$ illuminated slits.
	Directly behind the grating (left), the contributions of all slits overlap, leading to the formation of the Talbot carpet. 
	This pattern starts to transition into the far-field pattern after $(N-1)\cdot L_T$ revivals.
	}
	\label{fig:Talbot_lin}
\end{figure}

To elucidate this matter, we compute the diffraction pattern behind $N$-slits of period $d$ that are illuminated by a plane wave, as shown in Suppl.~Fig~\ref{fig:Talbot_lin}. 
In the simulation we include the quadratic terms to capture the effects of the wave front curvature.
Immediately behind the nanomask, we see the appearance of a shadow image as also expected from geometrical ray optics. 
With increasing distance, this pattern becomes blurred but reappears at multiples of the \textit{Talbot distance} $L_T=d^2/\lambda$.\cite{Patorski1989}
Plotting many interference patterns as a function of distance $z$ behind the grating adds up to a \textit{Talbot carpet}, which is periodic in $x$ with period $d$ and in $z$ with period $L_T$.\cite{Berry2001,Case2009}
This set of near-field images can be easily observed in any undergraduate optics lab using a collimated laser beam, a diffraction grating and a webcam.\cite{Case2009} 
The extension of that idea to 3-grating near-field interferometry is particularly appealing for realizing coherence experiments with incoherent sources and is routinely used for \textit{Talbot-Lau imaging} in UV lithography,\cite{Patorski1989} X-ray imaging,\cite{Pfeiffer2006} and in high-mass matter-wave interferometry.\cite{Brezger2002,Gerlich_NatPhys3_711,Haslinger2013,Fein_NatPhys15_1242}
In the latter case the intensity patterns correspond to the probability density $|\psi|^2$ to find a particle at any specific location.

\begin{figure}[t!]
	\includegraphics[width=\linewidth]{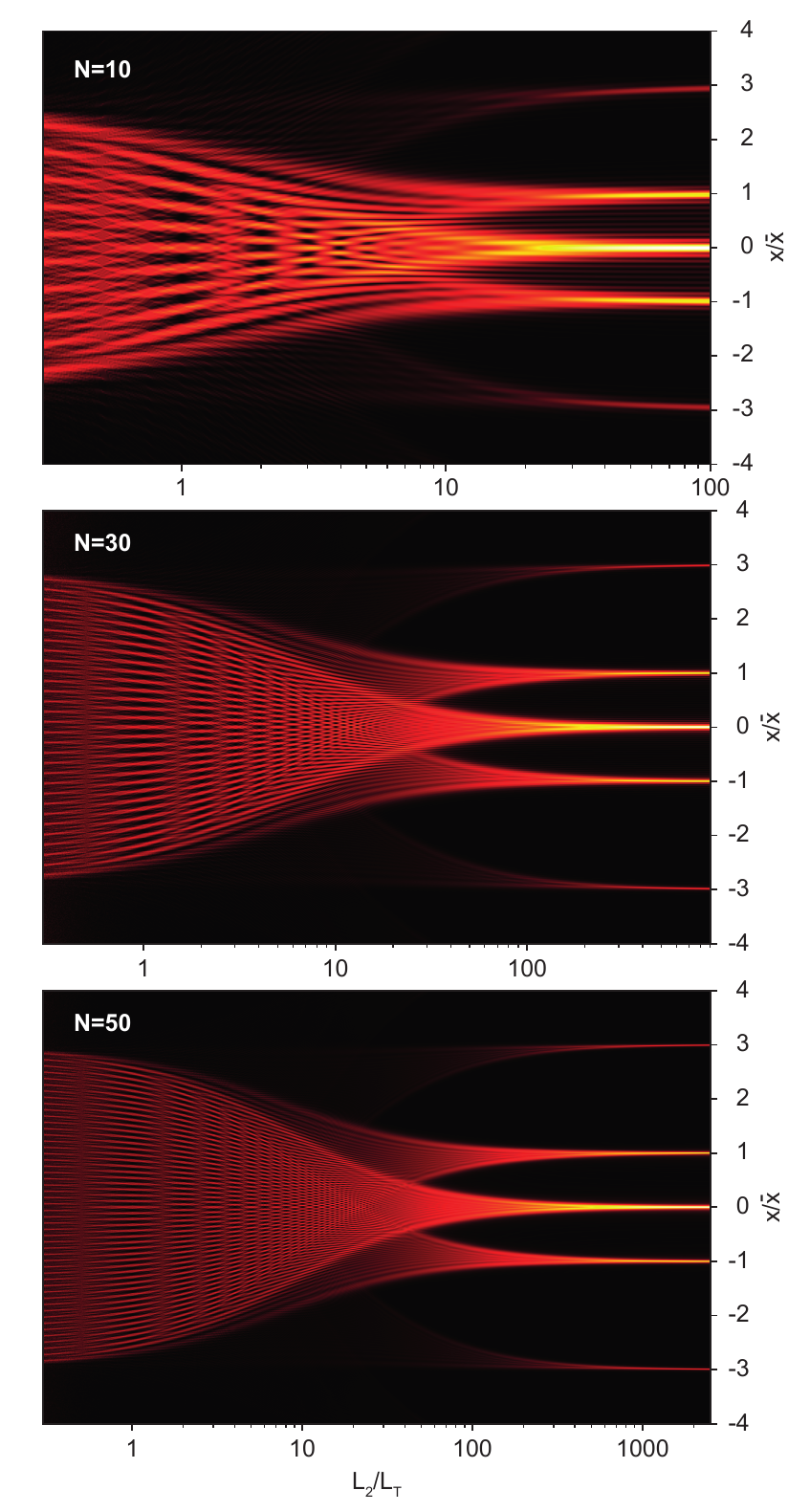}
	\caption{Transition from the near to the far field for $N=10$ (top), $N=30$ (middle), and $N=50$ (bottom) illuminated slits.
	Here, the distance behind the grating is depicted logarithmically in $L_2/L_T$, and $x$ is rescaled by $\bar{x}=L_2\lambda/d\cdot(1+NL_T/6L_2)$.
	This allows to capture both the near- and the far-field pattern.
	It illustrates that the distance after which the diffraction orders emerge depends on $N$.
	}
	\label{fig:Talbot}
\end{figure}

So why is the far field often a useful approximation even if the wave front curvature plays a role? 
The answer lies in the grating's width. 
Suppl.~Fig.~\ref{fig:Talbot} shows how the length of the Talbot carpet depends on the number $N$ of coherently illuminated slit: The self-images reappear $N-1$ times before they transition into the far-field pattern, whose diffraction orders are separated by $\sin \vartheta =n\cdot \lambda/d$. 
In our experiments with $\lambda_\mathrm{dB}=3\times 10^{-12}$~m and $d=100$~nm, the Talbot distance amounts to $L_T=3.33$~mm. 
At the position of the detector ($L_2=180~L_T$) the diffraction orders are already well-separated, even if our whole grating ($N=50$) is illuminated coherently.
While they still carry a fine structure, this is concealed by the collimation of the molecular beam.

\section{Fourier transform of the transmission function}
\label{sec:SupplFourier}

The far-field diffraction pattern of a thin grating can be found by applying Huygens' principle of superposing elementary waves.
In one dimension the intensity distribution $I$ at the detector is
\begin{equation}
I (x)\propto \left [\int_{w}  e^{ik_{\rm dB}x\xi/z} t(\xi) d\xi \right]^2
\label{eqn:Kirchhoff}
\end{equation}
with $k_{\rm dB}=2\pi/\lambda_{\rm dB}$.
Here the mask transmission function $t(\xi)$ depends on the transverse position $\xi$ inside the grating slits.
Linearizing the exponent (for large $z$) reduces the expression to the Fourier transform of the mask transmission function. 
For $N$ slits distributed over the grating width $w$, Eq.~(\ref{eqn:Kirchhoff}) results in Eq.~(1) of the main manuscript.
Thus, the diffraction minima and maxima are positioned where we expect them based on Huygens' wave argument. 
This rule holds waves both in optics and quantum mechanics.

\section{Coherence}
\label{sec:SupplCoherence}

Coherence is a quantitative statement about the existence of phase correlations between waves in space and time. 
Examples for highly coherent emitters are continuous narrow-band lasers or atom lasers emitting quantum degenerate atoms from Bose-Einstein condensates.\cite{Ketterle2002}  

Formally, coherence can be defined as a normalized correlation function $C\propto\langle \psi^*(\mathbf{r_1},t_1) \psi(\mathbf{r_2},t_2) \rangle $. 
Both for educational purposes and to describe advanced research experiments, it is often more intuitive to introduce transverse (spatial) and longitudinal (spectral) coherence separately. 
The correlation function can then be associated with two characteristic coherence lengths.
In the literature various approaches exist to define those coherence lengths, resulting in different prefactors.
We summarize these concepts here to clear-up the quantitative differences that students may find in the literature. 

\subsection{Transverse coherence}
Mathematically, transverse coherence is described by the van Cittert-Zernike theorem: 
it equates the coherence function behind an incoherently illuminated aperture with the functional form of the diffraction pattern of the same aperture under coherent illumination.\cite{Born_principles-of-optics}
For a rectangular slit this means that the coherence function $C(\theta)$ under monochromatic but spatially incoherent illumination is proportional to the intensity diffraction pattern $I(\theta)$ behind the same slit when illuminated with a plane wave: $C(\theta)\propto I(\theta) \propto \sin^2\beta/\beta^2$, with $\beta=(\pi s/\lambda)\sin\theta$. 

The coherence function $C(\theta)$ vanishes for $\beta=n\cdot \pi$, with $n\in \mathbb{N}$ and we can define the transverse coherence width $X_T$ as the distance between the two first-order minima. 
With $\pi = \beta=(\pi s /\lambda) \cdot\sin(\theta)\simeq (\pi s/\lambda) \cdot  X_T/2 L_1$ (in the small angle approximation), we thus get 
\begin{equation}
X_T \simeq \frac{2 L_1 \lambda_\mathrm{dB}}{s}
\label{eqn:transverse}
\end{equation}
\begin{figure}[tbh]
	\includegraphics[width=0.9\linewidth]{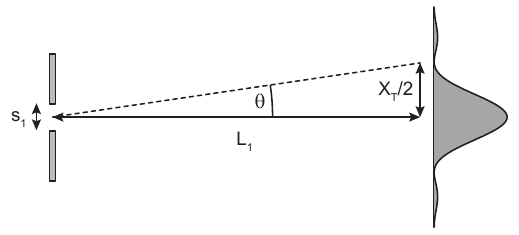}
	\caption{The transverse coherence function behind a source under spatially incoherent but monochromatic illumination is form-equivalent with the intensity diffraction pattern under spatially coherent illumination of the same aperture.} 	
	\label{fig:Koh}
\end{figure}

We find a pictorial interpretation of the van Cittert-Zernike theorem also by realizing that any wave can be decomposed into a complete set of plane waves. 
Each of them, incident under a different angle, will result in $\mathrm{sinc}(\theta)$- diffraction pattern on the screen. 
However, since these partial waves all impinge under different angles, they will cause diffraction patterns that are shifted with respect to each other. 
A wide source of uncorrelated emitters will therefore wash out the interference pattern entirely, while the underlying coherence is still associated with the individual diffracted wave.  

We can interpret transverse coherence also as a result of Heisenberg's position-momentum uncertainty relation. 
The source and any collimation slit along the molecular beam impose a transverse boundary condition on the position of the molecules. 
This is associated with an inherent indefiniteness $\sigma_p$ of the transverse momentum $p_x$.\cite{Shull_PhysRev179_752,Nairz_PRA65_032109}
The inequality
\begin{equation}
\sigma_x\cdot\sigma_p\geq \hbar/2
\end{equation}
connects the product of the standard deviation of momentum $p_x$ and position $x$ for Gaussian states.
However, in our single-slit diffraction experiment the wave is delimited by a rectangular transmission function of width $\Delta x=s$, whose Fourier transform defines the far-field momentum pattern: 
$I(\theta)\propto\sin^2(\beta)/(\beta)^2$ (see Suppl.~Fig.~\ref{fig:Koh}). 
The half-width of $I(\theta)=0.5$ is found at $\sin\theta=0.444\,\lambda_{\rm dB}/s$. 
If we identify this angle with the half-width of the transverse momentum uncertainty $\Delta p_x/2$, we find $\theta  \simeq \Delta p_x/p_z \simeq  0.89\,\lambda_{\rm dB}/s$ and thus
\begin{equation}
 \Delta x \cdot \Delta p_x  \simeq 0.89\,h~.
\label{Heisenberg}
\end{equation}
If we define the full width at half maximum (FWHM) of the quantum uncertainty as the 
intrinsic transverse coherence width, we see how its width in momentum translates into position:\cite{Nairz_PRA65_032109,Shull_PhysRev179_752}
\begin{equation}
     X_T\simeq \Delta v_x t  =  \frac{\Delta p_x L_1}{p_z} \simeq  \frac{0.89\,  L_1 \lambda_\mathrm{dB}}{s}~.
\end{equation}
The exact value of the prefactor in Eq.~(\ref{eqn:transverse}) and Eq.~(\ref{Heisenberg}) depends on the experiment and the threshold we define for coherence 'to be reduced'. 
This could be a reduction in contrast by a factor of two, by $1/e$,  $1/e^2$ or down to the first zero of the coherence function.
Equation~\ref{eqn:transverse} has proven useful as a rule-of-thumb in many real-world molecular diffraction experiments for the transverse coherence at a distance $L_1$ behind a slit of width $s$.

\subsection{Longitudinal coherence}
\begin{figure}[tbh]
	\includegraphics[width=0.8\linewidth]{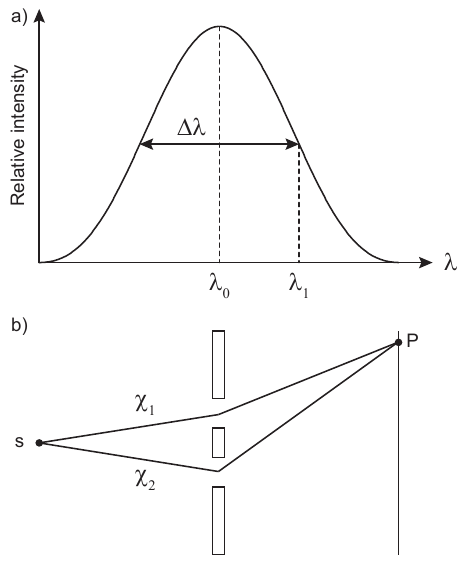}
	\caption{We consider a distribution of wavelengths $\lambda$ with a FWHM $\Delta \lambda=2(\lambda_1-\lambda_0)$ (a).
    The resulting longitudinal coherence length determines whether diffraction orders can be resolved at a point $P$ on the detection screen (b).}
    \label{fig:Verteilung}
\end{figure}

To define the longitudinal coherence length $X_L$, we consider a point source emitting a spectrum centered at $\lambda_0$ with a FWHM $\Delta \lambda=2(\lambda_1-\lambda_0)$. 
The emitted wavelets traverse a double-slit in a superposition of the upper and the lower path, as shown in Suppl.~Fig.~\ref{fig:Verteilung}.
One can define the coherence length as the shortest path-length difference between both arms from $s$ to $P$ for which an integer multiple of $\lambda_0$ (constructive interference) coincides with an odd-integer multiple of $\lambda_1/2$  (destructive interference):\cite{Demtroeder2008a}
\begin{equation}
X_L\simeq \frac{\lambda_0^2}{\Delta\lambda}\simeq \lambda_0\frac{v}{\Delta v} .
\label{eqn:longcoh}
\end{equation}
For matter-waves often the evolution of Gaussian wave packets is considered to establish a connection between the quantum wave description of matter and the notion of localized particles:\cite{Adams_PhysRep240_143, Tino2014}
\begin{equation}
\psi(\mathbf{r},t)\propto \int dk \, \exp^{-(k-k_0)^2/2\sigma^2_k}\exp^{i(\mathbf{k \cdot r}-\omega t)}.
\label{eqn:Gaussian}
\end{equation}
Here, $k=2\pi/\lambda$ is the wave vector and $\sigma_k$ is the standard deviation of its distribution, which is associated with the width of the velocity distribution $\sigma_v$. 
If we evaluate the correlation function, we find the longitudinal coherence length 
\begin{equation}
X_L=\frac{1}{2\pi}\lambda_0\frac{v}{\sigma_v}.
\label{eqn:lcoh} 
\end{equation}
While this suggests a six times smaller value than the definition of Eq.~(\ref{eqn:longcoh}), it is based on a different definition of $\Delta \lambda$: 
Equation~\ref{eqn:lcoh} uses the standard deviation $\sigma_v$ while it is the FWHM of $\Delta v$ for Eq.~(\ref{eqn:longcoh}). 

In the present experiments the assumption of a Gaussian wave packet is inadequate as we are dealing with continuous molecular beams in a mixed thermal state. 
To evaluate which definition describes the experiment best, we compare them: 
For $\Delta v/v=\Delta \lambda/\lambda=10\%$, as typically realized in our experiments, Eq.~(\ref{eqn:lcoh}) predicts $X_L\simeq \lambda_{\rm db}$, suggesting that we should observe at maximum the first diffraction orders. 
However, we observe interference up to the sixth diffraction order (Fig.~6), in good agreement with Eq.~(\ref{eqn:longcoh}).
 
\section{Thermal beam properties}
\label{sec:beam}
Since the de Broglie wavelength, the coherence length and the signal strength depend on many molecular beam properties, we summarize here the essentials of molecular beam physics.\cite{Pauly2000}
Let the molecules be sublimed in an oven, which is often also referred to as a Knudsen cell. 
In the following, we treat the molecules as an ideal gas which usually provides a good approximation in the cases discussed here.

\subsubsection{Thermalized gas} The probability to find a molecule with speed $|v|$ is given by the Maxwell-Boltzmann distribution:
\begin{equation}
    \label{eqn:maxwell-boltzmann}
    f(v)_v dv=\frac{4}{v_\mathrm{mp}\sqrt{\pi}}\left( \frac{v}{v_{\mathrm{mp}}}\right)^2\exp\left(-\frac{v^2}{v^2_{\mathrm{mp}}}\right) dv
\end{equation}
with  $v_{\rm mp}=\sqrt{2k_{\rm B}T/m}$.
In this gas we can define the 
\begin{enumerate}
    \item Most probable velocity $v_{\rm mp}$: 
    \begin{equation}
        df(v)/dv=0  \rightarrow v_\mathrm{mp}
    \end{equation}
    \item Average velocity $\bar{v}$: 
    \begin{equation}
        \bar{v} = \int_0^\infty v f(v) dv
    \end{equation}
    \begin{equation}
        \rightarrow \bar{v} = \sqrt{8k_{\rm B}T/\pi m}
    \end{equation}
    \item Root mean square (rms) velocity:
    \begin{equation}
        \bar{v^2} = \int_0^\infty v^2 f(v)dv=3k_{\rm B} T / m
    \end{equation} 
    \begin{equation}
        \rightarrow \sqrt{\bar{v^2}}=\sqrt{3k_{\rm B}T/m}
    \end{equation}
\end{enumerate}
The full width at half maximum of the velocity distribution is then $\Delta v = 0.993 v_\mathrm{mp}\simeq v_\mathrm{mp}$.

\subsubsection{Molecular flux} 
The molecular flux is given by the product of the particle density, the velocity and its probability distribution function and is therefore more skewed towards higher velocities. With the particle density $n$ we find:
\begin{equation}
    \label{eqn:distribution-flux}
    \Phi(v) dv = 2 n \left ( \frac{v}{v_{\mathrm{mp}}}\right )^3 \exp\left(-\frac{v^2}{v^2_{\mathrm{mp}}}\right) dv .
\end{equation}

When having a thermal beam impinging a particle detector, it is also of importance to distinguish the particle's intrinsic velocity distribution as given in Eq.~(\ref{eqn:maxwell-boltzmann}) from the velocity distribution of the particles impinging the detector. The latter is proportional to the flux, since due to their speed faster particles collide with the detector more often than slow ones. With the correct normalization, one ends up with:
\begin{equation}
    f_{v}(v) dv = \frac{2}{v_{\mathrm{mp}}} \left ( \frac{v}{v_{\mathrm{mp}}}\right )^3 \exp\left(-\frac{v^2}{v^2_{\mathrm{mp}}}\right) dv .
\end{equation}
The velocity that corresponds to the highest molecular flux is then
$v_\mathrm{hf}=\sqrt{3k_{\rm B} T/m}$ , which is higher by a factor of $\sqrt{3/2}$ than the most probable velocity $v_{\mathrm{mp}}$.

\subsubsection{Lambert emitter } 
A thermal source is an isotropic emitter (Lambert emitter), whose apparent surface area depends on the angle of observation:
\begin{equation}
    A_\mathrm{eff}=A\cdot \cos \theta .
\end{equation}
With this and the Maxwell-Boltzmann distribution, one can find the forward directed angular flux, i.e. the number of molecules emitted from the surface area $A$ per unit solid angle and second:\cite{Pauly2000}
\begin{equation}
\frac{dI(\theta = 0)}{d\Omega}=8.4\times 10^{21} \cdot \frac{p_0 \text{[hPa]} \cdot A[\text{cm}^2] }{\sqrt{m[\text{u}] \cdot T[\text{K}]}} \text{molec. sr}^{-1}\text{s}^{-1} .
\end{equation}
Here, the saturated vapor pressure $p_0$ is measured in mbar and depends exponentially on temperature, following the Clausius Clapeyron equation 
\begin{equation}
    \log P (T) \,[10^5 \text{Pa}] = A-B/T,
\end{equation}
with $A=12.34$ and $B=11687 K$ for phthalocyanine.\cite{Basova2009}
These equations allow us to precisely predict the number of molecules in the forward beam.

\section{Suggested problems}
Since molecule diffraction is increasingly used in high schools as an example for the quantum wave-particle duality, we provide below a number of problems and solutions around molecular far-field diffraction, which may be used as exercises. As the diffraction angles are typically in the µrad-regime a  small-angle approximation $\sin\theta \simeq \theta\simeq \tan \theta$ is always justified in the following.

\subsection{Definition of experimental parameters}

\begin{enumerate}
    \item {\em Molecule:} Phthalocyanine (PcH$_2$, C$_{32}$H$_{18}$N$_8$) is a planar, non-polar 
    molecule with a mass of $m=514.54$~u and a diameter $D$ of $1.5$~nm in the molecular plane. 
    \item {\em Geometry:} The source has a slit width of $s_1=1.4~$µm and the molecular beam is collimated by a second slit $s_2$.
    These two slits are separated by $L_1=1$~m and the grating $G$ is so close to $s_2$ that we can use $L_1$ also as the distance between $s_1$ and $G$. 
    The distance between the grating and the detector is $L_2=1$~m.
    \item {\em Grating:}  The grating has a period $d=100$~nm, a slit opening of $s=50$~nm and a thickness of $T=10$~nm.
    The geometric open fraction is then $f=s/d=0.5$.
\end{enumerate}

\subsection{Problems for students}
\begin{enumerate}
    \item {\em De Broglie wavelength:} Calculate the de Broglie wavelength $\lambda_{\rm{dB}}$ of PcH$_2$ travelling at $v=250$~m/s and compare it to the wavelength of red light (600~nm) and to the diameter $D$ of the molecule.
    \item {\em Diffraction angle:} At which angle $\theta_{\rm diff}$ do you expect the first diffraction order for these molecules?
    \item {\em Collimation:} Consider $s_1$ to be a point source emitting PcH$_2$ at $v=250$~m/s. \\
    a) Calculate the diffraction angle $\theta_{\rm diff}$. \\
    b) Compute the slit width $s_2$ that is needed to ensure that the diffraction orders are not overlapping. \\
    c) How wide may $s_2$ be if $s_1=10$~µm?
    \item {\em Transmission:} Molecules are isotropically emitted from a thermal source. 
    Estimate the reduction in molecular flux due to the collimation slit $s_2$ and the grating structure.
    The slit $s_2$ is $20$~µm high and $5$~µm wide, and the grating including support structure has a transmissivity of $30 \%$.
    \item {\em Transverse coherence:} Compute the maximal source size $s_1$ that is compatible with the requirement that two grating slits are coherently illuminated by PcH$_2$ with $\lambda_{\rm dB}=3.1\times 10^{-12}$~m (see Eq.~(\ref{eqn:transverse})).
    \item {\em Longitudinal coherence:} Let the molecular beam have a velocity distribution, peaked at $v=250$\,m/s with a FWHM of $\Delta v=50$~m/s. 
    Compute the longitudinal coherence length $X_L$ of the beam and  express it in multiples of the central de Broglie wavelength.
    \item {\em Single slit diffraction:} In the absence of any grating  close the second collimation slit down to $s_2=70$~nm. 
    Which pattern do you expect on the screen further downstream for PcH$_2$ moving at $250$~m/s? 
    Compare the classical and quantum expectation for its full width at half maximum. 
    \item {\em Effective open fraction:} Consider a grating with an open fraction $f$ of $0.5$. Which interference orders are systematically suppressed on the screen and why is that so?
    Why are they not suppressed in diffraction of organic molecules at such a nanomechanical grating?
    \item {\em Velocity selection:} 
    Consider a beam of molecules with a velocity distribution $\Delta v$ diffracted at a nanomechanical grating. \\
    a) If at a given height of the detector all velocities contribute, which parameters define how many diffraction orders are visible? \\
    b) How does the situation change if we introduce a vertical slit right before the grating?
    \item {\em Earth's rotation:} A molecule interferometer is oriented with its beam flying from south to north. 
    What is the effect of the rotation of the Earth?
    Estimate the expected fringe shift for a latitude of 45$^\circ$ and a molecular velocity $v=250$~m/s. 
    Can you see this in the experiment?
    \item {\em Laser evaporation source:} A thin film of PcH$_2$ with a surface density of $\rho_{\rm surf}=8$ g/m$^2$ is prepared on a glass slide.
    A laser beam is tightly focused onto the molecules and evaporates them from an area with a diameter of 1~µm. 
    To replenish new material, the glass slide moves at a constant velocity of $v=1$~cm/s underneath the laser beam. 
    How many molecules contribute to the interference image per hour?
    Use the fraction of contributing molecules from the problem 2.4.
    \item {\em Thermal beam:} Let phthalocyanine molecules be sublimed in an oven at $T=900$\,K and treat them as an ideal gas. 
    The molecules can leave the oven through a small hole of area $A=0.1\times 0.1$\,mm$^2$.
    With the definitions given in Sec.~\ref{sec:beam}: \\
    a) Compute the most probable velocity in the beam as well as the associated de Broglie wavelength. \\
    b) Calculate the thermal coherence length. Will it be possible to see interference after passing a grating?
\end{enumerate}
 
\subsection{Solutions for students}
\begin{enumerate}
    \item {\em De Broglie wavelength:} 
    The de Broglie wavelength $\lambda_{\rm{dB}}=h/mv = 3.1\times 10^{-12}$~m is about $480$ times smaller than the molecule itself and about $1.9\times 10^5$ times smaller than the wavelength of red light ($600$~nm).
    \item  {\em Diffraction angle:} 
    The first-order diffraction angle is $\theta_{\rm diff}=\lambda_{\rm{dB}}/d \simeq 3.1\times 10^{-5}$\,rad. 
    This justifies that we treat all angles as small and paraxial.
    \item {\em Collimation:} 
    To separate the diffraction orders, the diffraction angle must exceed the collimation angle.
    This yields $s_2< 30~$µm for the diffraction angle of problem 2.2 and thus $s_2=30$~µm.
    For an extended source the collimation angle is given by $\theta_\mathrm{coll}=(s_1+s_2)/L_1$ (see Suppl.~Fig.~\ref{fig:Kollimation}).
    For $s_1=10$~µm the width of the collimation slit must be reduced to $s_2< 20~$µm.
\begin{figure}[h!]
	\includegraphics[width=0.9\linewidth]{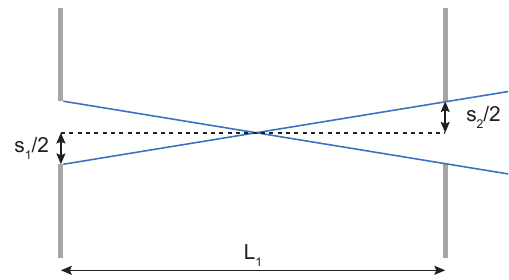}
	\caption{ The half-angle between source on the left and the collimation slit on the right is $(s_1+s_2)/2L$.}
	\label{fig:Kollimation}
\end{figure}
    \item {\em Transmission:} We can estimated the effective solid angle filled by $s_2$ as seen from the source. 
    In combination with the fraction of transmitted molecules we get:
    \begin{equation}
    0.3\cdot \frac{\Omega}{2\pi} =0.3\cdot \frac{5~\text{µm}\cdot 20~\text{µm}}{2\pi\cdot L_1^2} =4.8\times 10^{-12}. \nonumber
    \end{equation}
    Less than one out of 100 billion emitted molecules contribute to the final interference pattern.
    \item {\em Transverse coherence:}  If we take $X_T=2\lambda_\mathrm{dB} L_1/s_1> 150$~nm (two slits and a grating bar), we demand that $s_1<43$~µm. 
    Typically, the source is chosen to be smaller to ensure that the diffraction orders are not overlapping, see problem 2.3.
    \item {\em Longitudinal coherence:} A practical definition for longitudinal coherence is 
    \begin{equation}
    X_L\simeq \lambda^2_{\mathrm{dB}}/ \Delta \lambda_{\rm{dB}} \simeq \lambda_{\mathrm{dB}} \cdot v/ \Delta v=5\,\lambda_{\rm{dB}} .\nonumber
    \end{equation}
    Since with every diffraction order the path length difference across the grating grows by $\lambda_{\rm{dB}}$, we expect to see diffraction up to the fifth diffraction order.
    \item {\em Single-slit diffraction:}
    For a classical particle, we expect a shadow given by the collimation.
    The width of the signal at the detector is hence $x_{\rm class.}=(s_1+s_2)/L_1\cdot (L_1+L_2)=2.9$~µm.
    For a quantum particle, we have to obey Heisenberg's uncertainty relation for diffraction through a slit (see Eq.~(\ref{Heisenberg})):
    \begin{align}
        \Delta p_x&=m\cdot \Delta v_x\simeq \frac{0.89~h}{s_2} \nonumber \\
        &\rightarrow x_{\rm quant.}=\frac{\Delta v_x\cdot L_2}{v}\simeq\frac{0.89~h~L_2}{s_2\cdot m\cdot v}=39.4~\text{µm} \nonumber 
    \end{align}
    The FWHM of the signal is about 13 times wider than the pattern of a classical particle and depends on the molecular velocity, which is not the case for the geometrical shadow.
    Moreover, we expect to see the side lobes of the sinc-function, which are also not present for the shadow.
    \item {\em Effective open fraction:} 
    In far-field diffraction, the interference maxima of the multi-slit pattern are located at $\sin \theta = n\lambda/d$, with $n\in N$, while the single-slit minima are aligned along $\sin \theta_\mathrm{min} = n\lambda/s$, with $n\in N$. 
    With $d=2s$, all even orders are suppressed.
    For molecules $s_2$ is reduced by the van der Waals interaction to an effective slit width.
    As $d\neq 2s_{\rm eff}$, the even diffraction orders are not suppressed any more.
    \item {\em Velocity selection:} 
    Without any velocity selection, the visibility depends on the grating parameters ($s$ and $d$) and on $\Delta v/v$ as the latter defines the coherence length.
    Introducing a vertical slit before the grating restricts the range of molecular velocities that can contribute to the pattern at a certain height of the detector. 
    This changes $\Delta v$ and thus increases the longitudinal coherence length. 
    Thus, we can observe more diffraction orders.
    \item {\em Earth's rotation:} The Earth's rotation introduces an inertial acceleration on any massive object in free flight, the {\em Coriolis acceleration} $a_C=2 v \times \Omega_\mathrm{E}$, where $\Omega_\mathrm{E}=72\,\mu$rad$/s$ is the rotation rate of the Earth. 
    At a latitude of $\theta_L=45^\circ$, the acceleration acting on particles flying north is given by $a_C=2 v \Omega_\mathrm{E} \sin (\theta_L)=0.025\,$m/s$^2$, which causes a molecular beam shift on the detector of $\delta x=a_c (L_1+L_2)^2/2v^2= \Omega_\mathrm{E}\sin (\theta_L) (L_1+L_2)^2/v\simeq 0.8~$µm. 
    This is an order of magnitude smaller than the width of the collimated molecular beam and thus hardly detectable.
    \item {\em Laser evaporation source:}
     During the time $t=3600$~s the laser beam depletes the area $A=D \cdot v t=3.6\times 10^{-5}$~m$^2$.
     This corresponds to $N_{\rm evap}=\rho_{\rm surf}\cdot A \cdot N_A/m_{\rm PcH2}=3.37\times 10^{17}$ molecules with Avogadro's number $N_A$. 
     Using the fraction of molecules contributing to the diffraction pattern (exercise 3.4), we expect to find $N \simeq 1.6\times 10^6$ molecules in one hour.
    \item {\em Thermal beam:}
    a) The most probable velocity for PcH$_2$ at $900$~K is $v_{\rm mp} = \sqrt{2k_{\rm B}T/m} \simeq 171$~m/s, which results in a de Broglie wavelength of $\lambda_{\rm dB} = h/mv_{\rm mp} = h/\sqrt{2k_{\rm B}Tm} \simeq 4.5\times10^{-12}$~m.\\
    b) The thermal longitudinal coherence length can estimated by $X_L \simeq \lambda_{\rm dB}\cdot v_{\rm mp}/\Delta v$. 
    Using $\Delta v \simeq v_{\rm mp}$, we end up with $X_L \simeq \lambda_{\rm dB}$. In order to see the first interference maximum, it is necessary to have a longitudinal coherence length which covers at least the path length difference of $\lambda_{\rm dB}$. Since this condition is at the edge of being fulfilled, interference is possible, even though the first diffraction order will already start to appear washed out.
    
    For a more precise estimation of the longitudinal coherence length it would be necessary to take the de Broglie wavelength distribution of the particles impinging the detector. This can be calculated by transforming the velocity distribution given in Eq.~(\ref{eqn:distribution-flux}), using $\lambda_{\mathrm{dB}} \equiv h/mv_{\mathrm{mp}}$:
    \begin{eqnarray}
        f_\lambda(\lambda) d\lambda &=& f_v(v) dv = f_v(v(\lambda)) \frac{dv}{d\lambda} d\lambda \\
        f_\lambda(\lambda) d\lambda &\propto& \left(\frac{\lambda_{\mathrm{dB}}}{\lambda}\right)^3 \exp\left(-\frac{\lambda_{\mathrm{dB}}^2}{\lambda^2}\right) \frac{h}{m\lambda^2} d\lambda
    \end{eqnarray}
    This distribution's maximum is located at $\lambda_{\mathrm{max}} = \sqrt{2/5} \cdot \lambda_{\mathrm{dB}}$, while its FWHM can be numerically calculated as $\Delta\lambda \simeq 0.51 \cdot \lambda_{\mathrm{dB}}$. These values then result in $X_L \simeq \lambda^2_{\mathrm{max}} / \Delta\lambda \simeq 0.78 \cdot \lambda_{\rm dB}$.
     
\end{enumerate}
 
\subsection{Advanced problems} 

\begin{enumerate}
    \item {\em Velocity selection:} To improve longitudinal coherence, the spectrum of visible light can for instance be narrowed down by dichroic filters, prisms, monochromators, or cavities. 
    What tools and techniques could be used to narrow the de Broglie spectrum of atomic or molecular matter-waves?
    \item {\em Earth's gravity:} As soon as the height of a thermal beam is restricted by a horizontal slit right before the grating, the diffraction pattern displays parabolic rather than straight interference lines -- in contrast to the case of light.
    Explain this observation. 
    \item {\em Vibrations:} Compute for phthalocyanine the number of vibrational modes as well as their average population and estimate the total internal vibrational energy in electron volts, under the assumption that every mode is thermalized at $T=600$~K. 
    Assume for simplicity that each mode can be approximated as an harmonic oscillator with an energetic spacing of $E_v=0.1\,$eV.
    \item {\em Single molecule interference:} Taking the flux of molecules from problems 2.4 and 2.11, compute the density of the molecules at the location of the grating and compare their average distance with those of intramolecular forces you know. 
    Which information about the grating can be extracted from the diffraction pattern? 
    What would be required to enable for two-molecule interference? 
    \item {\em Thermal decoherence:} Why is the high internal excitation of the molecules in problem 4.3 still compatible with high-contrast de Broglie interference? 
    When do you expect a significant influence on the interference pattern?
    \item {\em Rotational excitation:} Based on the moment of inertia of PcH$_2$ in the molecular plane ($I=9.46\times 10^{-44}$~kg~m$^2$), compute the rotational energy, the most probably rotational quantum number and the classical angular frequency $\omega_\mathbf{rot}$ at $T=600\,K$.
    For simplicity, treat the molecule as a linear rotor.
    \item {\em Interactions with external fields:} What processes can occur if you expose PcH$_2$, having a polarizability $\alpha$, to a) a constant electric E-field, b) constant $\mathbf{E}\nabla \mathbf{E}$ field? c) What is required to deflect polar molecule with a static dipole moment $\vec{d}$?
    \item {\em Grating transit:} Compare the transit time of the PcH$_2$ molecule with $v=250$~m/s through the 10\,nm thin grating with the most probable rotation period.
    \item {\em Internal clock and which-path information:} Argue: A rotating polar molecule could be regarded as the hand of a moving clock. Why can you still see high-contrast interference, even though n-th order diffraction requires a path length difference of $n \lambda_\mathrm{dB}$?
\end{enumerate}

\subsection{Solutions for the advanced problems}
\begin{enumerate}
    \item {\em Velocity selection:} The de Broglie wave spectrum can be narrowed by reducing the velocity spread. \\
    a) For atoms this can be very efficiently done using a combination of laser cooling techniques.\cite{CohenTannoudji1992}
    These led to the realization of Bose-Einstein condensed atomic ensembles where almost all atoms occupy the same quantum state.\cite{Cornell2002,Ketterle2002}  \\
    b) For general atomic or molecular beams velocity selection can be realized by gravitational free-fall, or by combining a pulsed source with some time-of-flight (ToF) measurement. 
    ToF measurements can be realized by mechanically moving slits (rotating disk velocity selectors or random choppers\cite{Scoles1988}) or by exploiting the high time resolution of secondary electron multipliers.\cite{Carnal1991,Carnal1995}
    
    \item{\em Earth's gravity:}
    Matter-waves of different de Broglie wavelengths travel at different speed even in vacuum, because $\lambda_\mathrm{dB}=h/mv$.
    The lateral position of the $n$-th interference order is determined by $X_n=n\cdot L_2 \cdot\theta_{\rm diff}  \simeq  n L_2 \lambda_\mathrm{dB}/d$. 
    Thus, by measuring the spacing of the diffraction orders on the screen, we can assign the velocity $v=n L_2 h/mX_nd$ of the molecules contributing: Slow molecules display a wider diffraction pattern. 
    Their vertical arrival point on the screen is determined gravitational free-fall $Y= -g t^2/2= g L_2^2/2v^2$. This yields the vertical parabola 
    \begin{equation}
        Y= -\frac{g  m^2 d^2}{2 n^2  h^2} X^2. 
    \end{equation}
    \item {\em Vibrations:} Any non-linear molecule consisting of $N$ atoms has $3$ translational and $3$ rotational degrees of freedom, as well as $3N-6$ vibrational modes. As phthalocyanine has 58 atoms, it has 168 vibrational degrees of freedom.
    
    Let every of all 168 vibrational modes be thermalized with $E_T = k_{\rm B} T/2$.
    The total vibrational energy is then $E_\mathrm{int} = 168 \, E_T\simeq 4.3$~eV.

    The expectation value for the rotational excitation is determined by Boltzmann's law: $\bar{n} = \exp(-E_v/k_{\rm B} T) = 0.144$.
    \item {\em Single molecule interference:} 
    $1.6\times 10^{6}$ molecules per hour makes $444$ molecules per second.
    Assuming all move with the most probable velocity of $250$~m/s, the average distance between them is $0.56$~m.
    This is much larger than the distance of a hydrogen bond ($100-200$~pm) or of a van der Waals interaction ($<600$~nm).
    Thus, there is no interaction between the molecules during the experiment.
    
    The information the molecules acquire are the effective slit width connected to the strength of the van der Waals interactions and the period of the grating. 
    These can be extracted from the single-slit diffraction envelope and the diffraction angles of the multi-slit interference.
    
    Interference requires indistinguishability in all degrees of freedom.
    For neutrons this requires 'only' to adjust position, momentum, energy, and spin, which is nonetheless experimentally hard.\cite{Iannuzzi_PhysRevLett96_080402}
    For atoms, this requirement seems to become much more demanding  -- because atoms have infinitely many internal states. 
    However, atoms can be extremely well controlled by laser light, both in their motion and in their internal states. 
    It has thus become almost routine in advanced research labs to prepare large ensembles of atoms at nK temperatures and all of them in only a single internal state.
    In these Bose-Einstein condensates (BEC), all atoms are indistinguishable. 
    They all contribute to the same interference pattern. 
    However, the de Broglie wavelength is always that of a single atom alone and removing one atom from the ensemble does not destroy the BEC. 
    To first order, even in these ultra-cold quantum degenerate gases one probes single particle interference. 
    However, higher order coherence has also been studied, meanwhile.\cite{Jeltes_Nature445_402}
    In all quantum interference experiments with complex molecules so far, every single molecule interfered with itself. 
    Multi-particle interference will still be precluded until sometime in the future when it may become possible to prepare even complex molecules in just a single and well-defined internal state.
    \item {\em Thermal decoherence:} Molecules can be considered their own internal heat bath: many dozens or even thousands of vibrational modes can couple to each other and also in our experiments as described in the main text, phthalocyanine is vibrating and rotating fast. 
    This might lead to relaxation via photon emission. 
    Photon emission entails three complementary processes: Firstly, the molecule provides some information about its position.
    From Bohr's complementary principle we know that you cannot have particle character (position) and wave character (interference) at the same time. 
    Secondly, we can follow Heisenberg's microscope argument for motivating his position-momentum uncertainty relation: if we want to resolve a very small distance, we need light of very short wavelength, incident from the broadest possible angle. 
    This means that a position measurement will add a momentum perturbation. Thirdly, we can argue that the emission of a photon entangles the molecule quantum mechanically with the environment. It turns out, that these three pictures are all quantitatively equivalent. 
    
    For undergraduate teaching, Heisenberg's microscope argument is often the most intuitive of all three. 
    If a single photon cannot resolve the different paths the molecule takes through the grating, scattering more photons can actually provide more optical information: The barycenter of a point spread function can be measured ever more precisely, the more photons are scattered. 
    In Heisenberg's words, this will however, induce a momentum diffusion, which after scattering of many photons will destroy the interference.
    
    Thermal decoherence has actually been observed in molecule interference experiments.\cite{Hackermueller2004}
    Note, that in complex molecules, rotational emission typically occurs in the microwave region, while vibrational photons have usually a wavelength between $3-20$~m, depending on the bond strengths. 
    Following Abbe's diffraction and microscopy theory, a single photon of even $3$~µm can at best resolve a path separation of $1.5$~µm in high vacuum. The emission or  scattering of a single infrared or microwave photon hence cannot destroy the interference pattern. 
    \item {\em Rotational excitation:}
    We can approximate PcH$_2$ as a oblate symmetric rotor with the following moments of inertia $I_A=I_B\simeq 9.46\times 10^{-44}$~kg~m$^2$ and $I_C\simeq 1.83\times 10^{-43}$~kg~m$^2$. 
    The respective rotational constants in Joule ($\hbar^2/2I$) are $A=B=5.88\times 10^{-26}$~J and $C=3.04\times 10^{-6}$~J.
    As we are interested in the most probable rotational quantum number $J_{\rm mp}$, we can set $K=0$. 
    Hence, $E_{\rm rot}=BJ(J+1)+(B-C)K^2$ simplifies to the expression for the linear rotor $E_{\rm rot}=BJ(J+1)$ as stated in the problem.
    
    The thermal distribution is given by 
\begin{equation}
    \frac{N_n}{N_0}=g_n\cdot \exp(-\frac{E_n-E_0}{k_{\rm B}T})
\end{equation}    
with the degeneracy $g_n=2J+1$.
Setting the derivative with respect to $J$ to zero results in
\begin{equation}
    J_{\rm mp}=\sqrt{\frac{k_{\rm B}T}{2B}}-\frac{1}{2}\simeq 265
\end{equation}
for $T=600$~K.

With $E_{\rm rot}=I_B\omega_{\rm rot}^2/2=k_{\rm B}T$, we get $\omega_{\rm rot}=\sqrt{2~k_{\rm B}T/I_B}=2\pi\cdot 67$~GHz.

The kinetic energy for rotation is $E_{\rm rot}=k_{\rm B}T=8.28\times 10^{-21}$~J ($\simeq 0.05$~eV).
    \item {\em Interactions with external fields:} 
    The response of a particle to an electric field $\mathbf{E}$ is governed by the Stark effect.
    For a non-polar but polarizable particle it leads to a shift of the electron density resulting in an induced dipole moment $\vec{d}_{\rm ind.}=\alpha \mathbf{E}$.
    The energy of this particle in the field is then $U=\vec{d}_{\rm ind.}\mathbf{E}$.
    The force acting on the particle is constant, if $\mathbf{E}\nabla \mathbf{E}$ is constant.
    
    A polar particle already has a dipole moment.
    Hence, only a $\nabla \mathbf{E}$-field is required to deflect the molecule.
    However, the interaction depends on the orientation of the dipole moment in the electric field, leading to an orientation-dependent force.
    This can be observed when diffracting polar molecules at charged gratings.
    As we sample over a huge number of molecular orientations, all patterns are shifted with respect to each other depending on the electric field.
    This can completely obscure the interference pattern.\cite{Knobloch_FortschrPhys65_1600025}

    \item {\em Grating transit:}
    The transit time for $v=250$~m/s through a $10$~nm thick grating is $4\times 10^{-11}$~s. 
    When we compare this to the duration of one rotation $t_{\rm rot}=2\pi/\omega_{\rm rot}$ for $J_{\rm mp}$ from problem 4.6, we get
    \begin{equation}
        \frac{t_{\rm rot}}{t_{\rm trans}}=\frac{1.5\times 10^{-11}}{2.4\times 10^{-12}}\simeq 6 .
    \end{equation}
    Thus, the molecules rotate several times during the transit.

    \item {\em Internal clock:}
    The dipole moment is fixed in the molecular frame and hence rotates with the molecule. 
    The contrast is reduced, when the gain of the clock along the different trajectories is significant.\cite{Arndt_Science349_1168}
    During one rotational period, a molecule moving at $v=250$~m/s travels $t_{\rm rot}\cdot v = 2\pi/\omega_{\rm rot} \cdot v = 3.75$~nm.
    This is $\approx 1200$ times longer than $\lambda_{\rm dB}$ at this velocity. Hence, the orientation remains virtually identical for the number of diffraction orders we are observing (up to 10).
    
    In general, decoherence occurs when the particle scatters or emits photons, collides with other particles, or leaves other traces in the environment, which allow to reconstruct which path was taken.
    The interaction of polar molecules with an electric field is conservative just as the van der Waals interaction.
    Following this argumentation, we expect to see a fully coherent diffraction pattern.
\end{enumerate}

\end{document}